\documentclass{ws-procs9x6}

\begin{document}
\title{Detuned Randall-Sundrum model:\\
Radion Stabilization\\ and \\ Supersymmetry Breaking
\footnote{\uppercase{T}alk presented by \uppercase{M. R.} at {\it
\uppercase{SUSY} 2003: \uppercase{S}upersymmetry in the
\uppercase{D}esert}\/, held at the \uppercase{U}niversity of
\uppercase{A}rizona, \uppercase{T}ucson, \uppercase{AZ},
\uppercase{J}une 5-10, 2003. \uppercase{T}o appear in the
\uppercase{P}roceedings.}}

\author{Michele Redi}

\address{Department of Physics and Astronomy\\
Johns Hopkins University\\
Baltimore, MD 21218 USA\\
E-mail: redi@pha.jhu.edu}


\maketitle

\abstracts{In this talk I describe the low energy effective theory
of the (supersymmetric) Randall-Sundrum scenario with arbitrary
brane tensions. The distance between the branes is stabilized, at
the classical level, by a potential for the radion field. In the
supersymmetric case, supersymmetry can be broken by a VEV for the
fifth-component of the graviphoton.}

\section{Introduction}
Warped compactifications offer a completely new perspective on the
hierarchy problem. In the Randall-Sundrum (RS) scenario
[\refcite{rs}], five-dimensional AdS space is compactified on an
orbifold $S^1/\mathbb{Z}_2$ with two opposite tension branes
located at the orbifold fixed points. The ratio between the Planck
mass and the electro-weak scale can be explained by the
gravitational red-shift of the metric along the fifth-dimension.
The hierarchy problem is rephrased in terms of the distance
between the branes which is a modulus of the compactification. A
stabilization mechanism is necessary.

In this talk I will consider a generalized version of the RS model
with ``detuned'' brane tensions [\refcite{kaloper}]. An intriguing
new feature of this scenario is the fact that the brane tensions
fix the distance between the branes in the vacuum. I will present
the low energy effective action for the radion for arbitrary
tensions. I derive the results in the supersymmetric version of
the model, where the low energy dynamics is controlled by an $N=1$
supersymmetric $\sigma-$model. The low energy effective action
includes a potential for the radion which stabilizes the size of
the extra-dimension. I will also argue that in the detuned
scenario supersymmetry can be broken spontaneously by a
non-trivial Wilson line for the graviphoton. The effective theory
in the non-supersymmetric scenario can be easily obtained from the
supersymmetric case. The material presented here is based on the
papers [\refcite{graviphoton}],[\refcite{radion}] in collaboration
with J. Bagger.

\section{The detuned Randall-Sundrum model}

The supersymmetric version of the model was constructed by Bagger
and Belyaev [\refcite{bagger2}]. The action is given by minimal
$5D$ supergravity supplemented by brane actions. Neglecting the
Chern-Simons term, the bosonic part of the action is simply
\begin{eqnarray}
S_{\rm bulk}&=&-\frac T {6 k}\int d^4x\int^\pi_{-\pi}d\phi
\sqrt{-G}\Big(\frac 1 2 R-6 k^2+\frac 1 4 F^{MN}F_{MN}\Big)\nonumber\\
S_{\rm brane}&=& -T_0 \int d^4x \sqrt{-g_0}-T_\pi \int d^4x
\sqrt{-g_\pi}, \label{action}
\end{eqnarray}
where $F_{MN}$ is the field strength of a $U(1)$ gauge field,
$B_M$, called the graviphoton. I work in the orbifold covering
space with the branes located at $\phi=0$ and $\phi=\pi$.
Supersymmetry requires the tensions to satisfy the bound
$T_{0,\pi}\le T$ [\refcite{bagger2}]. The full bulk-plus-brane
theory is invariant under five-dimensional $N=2$ supersymmetry in
the bulk, restricted to four-dimensional $N=1$ supersymmetry on
the branes. This guarantees that the effective theory is $N=1$
supersymmetric.

For generic tensions satisfying the bound above, the ground state
metric is four-dimensional AdS space warped along the
fifth-dimension,
\begin{equation}
ds^2=F(\phi)^2 g_{mn}dx^m dx^n+r_0^2 d\phi^2.
\end{equation}
Here the metric $g_{mn}$ is AdS$_4$ with radius $L$ and the warp
factor is,
\begin{equation}
F(\phi)=e^{-k r_0 |\phi |}+\frac 1 {4 k^2 L^2}e^{k r_0 |\phi|}.
\label{warpfactor}
\end{equation}
The radius $L$ is related to the tensions,
\begin{equation}
\frac 1 {4 k^2 L^2}=\frac {T-T_0} {T+T_0}.
\end{equation}
In contrast to the original RS scenario (which corresponds to the
choice $T_0=-T_\pi=T$), the radius $r_0$ of the extra-dimension is
fixed,
\begin{equation}
2\pi k r_0= \log \frac {(T+T_0)(T+T_\pi)} {(T-T_0) (T-T_\pi)}.
\label{radius}
\end{equation}
However the VEV of $B_5$ is not determined; it is the only modulus of the
compactification.

\section{Supersymmetric effective action}

The general form of the effective action is determined by
the symmetries of the five-dimensional theory up to four free
constants.

The bosonic low energy effective action includes the fluctuations
of the four-dimensional metric $g_{mn}$, together with the light
modes of $G_{55}$ and $B_5$. The scalar associated with $G_{55}$
can be identified as the proper distance between the branes, the
radion field. The other scalar is obviously the zero mode of
$B_5$. In the supersymmetric effective theory the two scalars join
with the fifth-component of the gravitino to form a chiral
multiplet. The zero mode of $B_5$ must be massless. The
Kaluza-Klein reduction fixes the mass of the radion to be $4/L^2$.
This is precisely the value of the mass required by the
representations of supersymmetry in AdS$_4$.

The effective action is $N=1$ supersymmetric, so it is determined
by a K\"ahler potential $K$ and a superpotential $P$. The bosonic
part of the action (setting $M_4=1$) takes the form
\begin{equation}
S_{\rm eff}=- \int d^4x \sqrt{-g} \Big[\frac{1}{2} R+K_{\tau
\overline{\tau}}g^{mn}\partial_m \tau \partial_n \overline{\tau} + e^{K}
(K^{\tau\overline{\tau}}D_\tau P {D_{\overline \tau} \overline{P}}- 3
P\overline{P})\Big],
\label{susyaction}
\end{equation}
where $\tau$ is the lowest component of the radion superfield, and
$D_\tau P=\partial_\tau P+ K_\tau P$. It can be shown that
$\tau=r+i b$ where $r$ is the radion field and $b$ is the zero
mode of $B_5$.

To determine $K$ and $P$, one can observe that the bosonic part of
the action (\ref{action}) is invariant under a shift of $B_5$.
This implies that, up to a K\"ahler transformation, $K$ is a
function of $\tau+\overline{\tau}$ (it does not contain $b$). By
the same argument, the potential in (\ref{susyaction}) is also a
function of $\tau+\overline{\tau}$. Since the superpotential is a
holomorphic function of $\tau$, this condition imply an infinite
number of constraints on $K$ and $P$. We found that, in an AdS$_4$
ground state, the most general solution of these constraints
is\footnote{This holds for $m_r^2=4/L^2$}
\begin{eqnarray}
K(\tau,\overline{\tau})&=&-3 \log[1-c^2\,
e^{-a (\tau+\overline{\tau})}]\nonumber\\[2mm]
P(\tau)&=&p_1+p_2\, e^{-3 a  \tau}, \label{kahler}
\end{eqnarray}
where $p_1$, $p_2 \in \mathbb{C}$ and $c$, $a \in \mathbb{R}$ are
undetermined constants.

With a simple change of variables one can recognize that this is
the K\"ahler potential of no-scale supergravity. In fact the
superpotential is the generalization to AdS$_4$ of the constant
superpotential of ordinary no-scale supergravity
[\refcite{radion}].

\section{Results}

The unknown constants in (\ref{kahler}) can be determined performing
the Kaluza-Klein reduction of the bosonic fields. This requires a
careful treatment of the tadpoles of the light fields with the
heavy fields [\refcite{radion}]. One finds that the K\"ahler potential
and the superpotential for the radion are given by
\begin{eqnarray}
K(\tau,\overline{\tau})&=&-3 \log\Big[1-e^{- \pi k(\tau+\overline{\tau})}\Big]
\nonumber
\\[2mm]
P(\tau)&=&\frac {k}{L} \sqrt{ \frac{6}{T}} \left(1 -
e^{\pi k r_0} e^{-3 \pi k \tau}\right). \label{summary}
\end{eqnarray}
The bosonic part of the action is then
\begin{equation}
S_{\rm eff}=- \int d^4x \Big[ \frac{M_4^2}{2}R  + 3 k^2 \pi^2 \,
M_4^2 \, \frac{e^{-k\pi (\tau+\overline{\tau})}} {(1-e^{-k\pi
(\tau+\overline{\tau})})^2}
 g^{mn}\partial_m \tau \partial_n \overline{\tau}+ V(\tau,\overline{\tau}) \Big],
\label{susyeffective}
\end{equation}
where I have inserted the four-dimensional Planck mass $M_4$. The
scalar potential is
\begin{equation}
V(\tau,\overline{\tau}) = -\frac {3\,M_4^2\,(1- e^{-2 k \pi
r_0})}{L^2}\left[ \frac{1-e^{-2 k \pi (\tau+
\overline{\tau}-r_0)}}{(1-e^{-k \pi (\tau+
\overline{\tau})})^2}\right]. \label{potential}
\end{equation}
As required, the potential is independent of $b$. The ground state
is AdS$_4$; the potential stabilizes the radius of the
extra-dimension (at $r=r_0$), while $b$ remains, at the classical
level, a modulus of the compactification.

The purely gravitational case can be obtained by setting $b=0$ in
(\ref{susyeffective}). For the case $T_{0,\pi}\ge T$ the $4D$
ground state is de-Sitter space and the $5D$ theory cannot be
supersymmetrized. The effective action can be still obtained from
the supersymmetric result (\ref{susyeffective}) replacing $L \to
iL$. The ground state is unstable in this case.

From the $5D$ point of view one can show that when $b$ has a VEV
(corresponding to a non-trivial Wilson line of the graviphoton),
the Killing spinor equations have no solutions; supersymmetry is
spontaneously broken [\refcite{graviphoton}] (see also
[\refcite{lalak}]. For a similar effect in flat space see
[\refcite{quiros}] and references therein). This effect however
vanishes when the tensions are tuned. In the effective theory,
unbroken supersymmetry requires that $D_\tau P$ vanish, when
evaluated at the minimum of the potential. It is easy to check
using (\ref{summary}) that supersymmetry is broken when $b\ne 2
n/(3 k)$, for $n$ integer. This mechanism of supersymmetry
breaking is the AdS$_4$ analog of the one in no-scale
supergravity.

\section*{Acknowledgments}
It is a pleasure to acknowledge the collaboration with J. Bagger.
This work was supported in part by the US National Science
Foundation, grant NSF-PHY-9970781.


\end{document}